# Quantitative characterization of spin-orbit torques in Pt/Co/Pt /Co/Ta/BTO heterostructure on the magnetization azimuthal angle dependence

Christian Engel, Sarjoosing Goolaup, Feilong Luo, and Wen Siang Lew[*]

*School of Physical and Mathematical Sciences, Nanyang Technological University,*

*21 Nanyang Link, Singapore 637371*

[*] wensiang@ntu.edu.sg

**Substantial understanding of spin-orbit interactions in heavy-metal (HM)/ferromagnet (FM) heterostructures is crucial in developing spin-orbit torque (SOT) spintronics devices utilizing spin Hall and Rashba effects. Though the study of SOT effective fields dependence on the out-of-plane magnetization angle has been relatively extensive, the understanding of in-plane magnetization angle dependence remains unknown. Here, we analytically propose a method to compute the SOT effective fields as a function of the in-plane magnetization angle using harmonic Hall technique in perpendicular magnetic anisotropy (PMA) structures. Two different samples with PMA, Pt/Co/Pt/Co/Ta/BaTiO$_3$ (BTO) test sample and Pt/Co/Pt/Co/Ta reference sample are studied using the derived formula. Our measurements reveal that only the dampinglike field of the test sample with BTO capping layer exhibits an in-plane magnetization angle dependence while no angular dependence is found in the reference sample. The presence of the BTO layer in the test sample, which gives rise to a Rashba effect at the interface, is ascribed as the source of the angular dependence of the the dampinglike field.**

The modulation of a ferromagnetic layer (FM) *via* an in-plane current in a heavy metal (HM)/FM bilayer is generally attributed to spin-orbit torque (SOT) effects [1, 2]. An understanding of the SOT effects is crucial for application devices using magnetization reversal [1–8], high-frequency oscillation [9–12], domain-wall- [13–15] and skyrmion motion [16, 17].

A charge current in the HM/FM wire directed along the *x*-direction generates dampinglike torque $\vec{\tau}_D = -H_D \vec{M} \times (\hat{m} \times \hat{y})$ and fieldlike torque $\vec{\tau}_F = -H_F \vec{M} \times \hat{y}$ on the FM layer, where $\hat{y}$ describes the direction of electron spin from spin-polarized current to the local magnetization and $\hat{z}$ the direction of the surface normal. The torques are characterized by the corresponding effective fields as dampinglike field $\vec{H}_D = H_D \hat{m} \times \hat{y}$ and fieldlike field $\vec{H}_F = H_F \hat{y}$ [18-21]. These spin-orbit torques are believed to be generated from the bulk spin Hall effect [1, 9, 22–24] (SHE) in the HM material with strong spin-orbit interaction and the Rashba effect [1, 25–27] at the interface with strong interfacial spin-orbit coupling. Yet, the identification of the dominant term remains difficult as both SHE and Rashba effects generate both torques with same expression [2, 23, 28].

Fan *et al.* [29] managed to identify the contributions of the Rashba effect from interface effects by inserting a copper layer at the interface to minimize the Rashba effect and distinguish effects from Rashba and from SHE in Pt/CoFeB bilayers. However, the ratio between dampinglike and fieldlike torques has been detected to vary between ~1 [30] and up to ~8 [31], which shows a rather complicated behavior. Ta-thickness dependence study using harmonic Hall measurement technique in Ta/CoFeB/MgO structure with perpendicular magnetic anisotropy (PMA) suggests a sign change of the fieldlike torque at small thicknesses of Ta (<1.5 nm) which is also nontrivial [32].

Theoretical predictions based on Boltzmann transport equation and diffusion theory suggest two mechanisms with different dependencies on the thickness of the HM layer [28], but they also predict fieldlike and dampinglike torques are independent on the rotation of magnetization normal to the film plane [20]. Garello *et al.* [33] showed a strong angular dependence on the magnetization polar angle with respect to the current direction which cannot be explained by the Boltzmann theory.

Another model based on the tight-binding theory predicts a strong angular dependence in spin-orbit torques with the Rashba spin-orbit coupling being comparable to the exchange coupling strength [34]. The origin of current-induced spin-orbit torques in HM/FM structures with PMA is still under debate. Hence, angular dependence measurements of SOTs are a powerful tool to deeply understand the underlying physics in HM/FM structures and to apply them into spin-orbit spintronics devices.

In this work, we investigate current induced effective fields in Pt/Co/Pt/Co/Ta/BaTiO$_3$ (BTO) and Pt/Co/Pt/Co/Ta reference sample using harmonic Hall measurement technique. We propose a method to measure the SOT effective fields for PMA materials as a function of the azimuthal angle $\varphi$ between magnetization and the current direction.

The dampinglike term is found to be ~17.0 Oe per $1 \times 10^{11}$ A/m$^2$ and the fieldlike term to be ~12.7 Oe per $1 \times 10^{11}$ A/m$^2$. The Pt/Co/Pt/Co/Ta reference sample without BTO shows no angular dependence of both SOT terms which is in agreement with the assumption of a weakly dependent azimuthal dependence of the SOT effective fields [30]. For the Pt/Co/Pt/Co/Ta/BTO test sample, an angular dependence of the dampinglike term is found with minima at 45° and 225° whereas no angular dependence is found for the fieldlike term. Rashba effect, which is caused from the BTO/Ta interface or from an intrinsic electric field within the weakly ferroelectric BTO, is ascribed as the source of the angular dependence of the dampinglike field. Also, large angular dependence of the effective anisotropy field $H_K$ is found for the Pt/Co/Pt/Co/Ta/BTO test sample whereas small variation in anisotropy field as a function of the azimuthal angle is found for the Pt/Co/Pt/Co/Ta reference sample. However, the relationship of the dampinglike field and the anisotropy field is still questionable.

# 1.    Analytical solution

In this section, we provide a theoretical model to extract the current induced effective fields in PMA magnetic structure with arbitrary magnetization orientation via Hall voltage modulation. Shown in Fig. 1 (a), is the schematic representation of a magnetic wire structure with Hall bars attached transverse to the long axis of the wire. For the magnetic wire, the transverse Hall voltage, $V_H$, typically contains contributions of the anomalous Hall effect (AHE), planar Hall effect (PHE) and spin Hall magnetoresistance (SMR) [32, 35-37] and can be written as:

$$V_H = \frac{1}{2} I \Delta R_A \cos\theta + \frac{1}{2} I \Delta R_{P\&S} \sin^2\theta \sin 2\varphi, \qquad (1)$$

where $\Delta R_A$ and $\Delta R_{P\&S}$ are Hall resistances which accounts for the resistance change due to AHE and PHE/SMR, respectively. $I$ is the constant current in the magnetic wire. $\theta$ and $\varphi$ are the polar and azimuthal angle of the magnetization vector $\vec{M}$ of the wire, as seen in Fig. 1 (a). The contributions from the PHE and SMR to the Hall voltage have identical symmetry and as such cannot be distinguished. For external fields applied in-plane with negligible out-of-plane contributions to the wire structure, the ordinary Hall effect can be neglected. The presence of an in-plane magnetic field, external or current induced, will lead to a deviation of the magnetization vector resulting in a change in Hall voltage. The stable polar and azimuthal angles of the magnetization vector $\left(\varphi, \theta\right)$ are determined by the applied field, $\vec{H}$. The current induced effective field, $\Delta\vec{H}$, modulates the magnetization vector via small deviations defined here as $\left(\Delta\varphi, \Delta\theta\right)$. For an alternating current (AC) source, the current induced harmonic modulation of the magnetization from the stable angle can be written as:

$$I \xrightarrow{ac} I_{ac} \sin \omega t ,$$

$$\Delta \varphi \xrightarrow{ac} \Delta \varphi \sin \omega t ,$$

$$\Delta \theta \xrightarrow{ac} \Delta \theta \sin \omega t .$$

where $\omega = 2\pi f$ is the angular frequency with frequency $f$ and $I_{ac}$ is the amplitude of the AC. For sufficiently small frequency, the change in magnetization angle will adiabatically trail the AC without any phase shift. The Hall voltage can be reformulated in terms of harmonic Hall voltages as:

$$V_H = V_0 + V_\omega \sin \omega t + V_{2\omega} \cos 2\omega t \qquad (2)$$

$V_0$ is a constant voltage, $V_\omega$ is the first harmonic Hall voltage and $V_{2\omega}$ the second harmonic Hall voltage. Substituting the induced harmonic magnetization modulation signals into Equation (1), and using Taylor expansion for small magnetization angle variations ($\Delta \varphi \ll 1$, $\Delta \theta \ll 1$), the constant Hall voltage is given as:

$$V_0 = \frac{1}{2}(B_\theta + B_\varphi) I_{ac} , \qquad (3)$$

where the terms $B_\theta$ and $B_\varphi$ are given by:

$$B_\theta = \frac{1}{2}\left[ -\Delta R_A \sin \theta + \Delta R_{P\&S} \sin 2\theta \sin 2\varphi \right] \Delta \theta , \qquad (4)$$

$$B_\varphi = \Delta R_{P\&S} \sin^2 \theta \cos 2\varphi \Delta \varphi . \qquad (5)$$

The first harmonic Hall voltage can be written as:

$$V_\omega = \frac{1}{2}\left[ \Delta R_A \cos \theta + \Delta R_{P\&S} \sin^2 \theta \sin 2\varphi \right] I_{ac} . \qquad (6)$$

The second harmonic Hall voltage is given by:

$$V_{2\omega} = -\frac{1}{2}(B_\theta + B_\varphi)I_{ac} . \tag{7}$$

The influence of the externally applied fields or current induced effective fields on the magnetization vector can be deduced from the total magnetic energy, $E_{Total}$, equation which is given as

$$E_{Total} = -\frac{1}{2}H_K M_S \cos^2\theta - \vec{M} \bullet \vec{H} , \tag{8}$$

where $H_K$ is the effective out-of-plane magnetic anisotropy field where $H_K = 2K_U / M_S - 4\pi M_S$ with $K_U$ being the uniaxial magnetic anisotropy energy, $M_S$ the saturation magnetization and $-4\pi M_S$ the demagnetizing field. The polar and azimuthal angles of the magnetization vector in the presence of an externally applied in-plane field, $\vec{H}$, can be derived by solving the stable energy derivatives $\partial E / \partial\varphi = 0$ and $\partial E / \partial\theta = 0$. The components of the external field with magnitude $H$ and applied along an azimuthal angle $\varphi_H$ acting along the $x$-axis and $y$-axis are given by $H_{X-ext} = H\cos\varphi_H$ and $H_{Y-ext} = H\sin\varphi_H$ respectively. The polar and azimuthal angles of the magnetization vector are then given as:

$$\varphi = \arctan\left(\frac{H_{Y-ext}}{H_{X-ext}}\right) , \tag{9}$$

$$\theta = \arcsin\left(\frac{H_{X-ext}\cos\varphi + H_{Y-ext}\sin\varphi}{H_K}\right) . \tag{10}$$

The current induced effective fields, $\Delta\vec{H}$, can be also decomposed into terms acting along the $x$-axis, $\Delta H_X$, and $y$-axis, $\Delta H_Y$, respectively. The current induced magnetization modulation can then be represented as

$$\Delta\varphi = \frac{\partial\varphi}{\partial\vec{H}}\Delta\vec{H} = \frac{\Delta H_Y H_{X-ext} - \Delta H_X H_{Y-ext}}{H^2_{X-ext} + H^2_{Y-ext}} \quad, \tag{11}$$

$$\Delta\theta = \frac{\partial\theta}{\partial\vec{H}}\Delta\vec{H} = \frac{\Delta H_X \cos\varphi + \Delta H_Y \sin\varphi}{H_K} \quad. \tag{12}$$

As the magnetization follows the external field adiabatically for low frequency of AC, the azimuthal angle of the magnetization can be set as $\varphi_H = \varphi$. By substituting Eqs. (9) and (10) into Equation (6), the first harmonic Hall voltage can be written as

$$V_\omega \approx \left[ \pm\frac{1}{2}\Delta R_A \left(1 - \frac{1}{2}\frac{H^2}{H_K^2}\right) + \frac{1}{2}\Delta R_{P\&S} \sin 2\varphi_H \frac{H^2}{H_K^2} \right] I_{ac} \quad. \tag{13}$$

The $\pm$ sign corresponds to the case where the initial magnetization vector is pointing along the $\pm z$-direction as the magnetization polar angles are $\theta$ for $+z$, $\pi - \theta$ for $-z$, and $\pm\Delta\theta$ for $\pm z$-direction respectively. Similarly, for the second harmonic Hall voltage, substituting Eqs. (9-12) into Equation (7) results in:

$$V_{2\omega} = \left( \frac{1}{4}\left[\pm\Delta R_A - 2\Delta R_P \sin 2\varphi_H\right]\left(\Delta H_X \cos\varphi_H + \Delta H_Y \sin\varphi_H\right) - \frac{1}{2}\Delta R_{P\&S} \cos 2\varphi_H \left(\Delta H_Y \cos\varphi_H - \Delta H_X \sin\varphi_H\right) \right)\frac{H}{H_K^2} I_{ac} \quad. \tag{14}$$

Taking the second derivative of Equation (13), $b_\omega = \frac{\partial^2 V_\omega}{\partial H^2}$, to remove the $H$-component from the equation results in:

$$b_\omega = \frac{\partial^2 V_\omega}{\partial H^2} \approx -\frac{I_{ac}}{2H_K^2}\left(\pm\Delta R_A - 2\Delta R_{P\&S} \sin 2\varphi_H\right). \tag{15}$$

Similarly as for Equation (15), by taking the first derivative with respect to $H$ in Equation (14), $b_{2\omega} = \frac{\partial V_{2\omega}}{\partial H}$ results in:

$$b_{2\omega} = \frac{\partial V_{2\omega}}{\partial H} = \left( \frac{1}{4} \left[ \pm \Delta R_A - 2\Delta R_P \sin 2\varphi_H \right] \left( \Delta H_X \cos\varphi_H + \Delta H_Y \sin\varphi_H \right) - \frac{1}{2} \Delta R_{P\&S} \cos 2\varphi_H \left( \Delta H_Y \cos\varphi_H - \Delta H_X \sin\varphi_H \right) \right) \frac{I_{ac}}{H_K^{\,2}} .(16)$$

The dependencies on $I_{ac}$ and $H_K$ can be removed by dividing Equation (16) by (15), $B_{\varphi_H} = \frac{b_{2\omega}}{b_{\omega}}$,

resulting in

$$B_{\varphi_H} = \alpha \Delta H_Y + \beta \Delta H_X , \qquad (17)$$

where the coefficients $\alpha$ and $\beta$ are given by

$$\alpha = -\frac{1}{2} \sin\varphi_H + \frac{\xi \cos 2\varphi_H}{\pm 1 - 2\xi \sin 2\varphi_H} \cos\varphi_H , \qquad (18)$$

$$\beta = -\frac{1}{2} \cos\varphi_H - \frac{\xi \cos 2\varphi_H}{\pm 1 - 2\xi \sin 2\varphi_H} \sin\varphi_H . \qquad (19)$$

$\xi$ is the ratio of P&S resistance ($\Delta R_{P\&S}$) to AHE resistance ($\Delta R_A$), $\xi = \Delta R_{P\&S} / \Delta R_A$. As seen from

Eqs. (18, 19), the coefficients $\alpha$ and $\beta$ are functions of the applied angle $\varphi_H \left( = \varphi \right)$ and the ratio $\xi$.

Given that $\Delta H_X$ and $\Delta H_Y$ are two independent variables, to extract their respective values from the measurement of $B_{\varphi_H}$, two independent measurements need to be carried out and prior knowledge of the coefficients $\alpha$ and $\beta$ are required.

The term $\alpha$ and $\beta$ are dependent on both the orientation of the applied field and the material parameter, $\xi$. Shown in Fig. 1 (b-d), are representative plots of $\alpha$ and $\beta$ as a function of the azimuthal angle of the external field with the ratio $\xi$ set to 0, 0.2 and 0.4 respectively. For the limiting case where $\xi = 0$, which corresponds to no effective PHE or SMR, $\alpha$ and $\beta$, exhibit a sinusoidal trend as a function of the azimuthal angle of the applied field. For ferromagnetic material with an effective PHE&SMR, where $\xi > 0$, characteristically different trends for $\alpha$ and $\beta$ are observed as function of the angle as seen in Fig. 1 (c) and (d). The trend is highly dependent on the magnitude of $\xi$. Irrespective of the value of $\xi$, the curves for $\alpha$ and $\beta$ intersect at azimuthal angle of 45° and 225°. For azimuthal angle of 0°, the magnitude of the coefficients $\alpha$ and $\beta$ are given by $\alpha = \xi$ and $\beta = -0.5$ respectively. Interestingly, at azimuthal angle of 90°, the values of the coefficient have swapped resulting in $\alpha = -0.5$ and $\beta = \xi$ which indicates a mirror symmetry at angle 45°. As such, the dependence of the coefficients can be mathematically written as:

$$\alpha(\xi, \varphi_H) = \beta'(\xi, \varphi'_H), \text{ where } \varphi'_H = 90° - \varphi_H \quad , \tag{20}$$

$$\beta(\xi, \varphi_H) = \alpha'(\xi, \varphi'_H), \text{ where } \varphi'_H = 90° - \varphi_H \quad . \tag{21}$$

The coefficients $\alpha'$ and $\beta'$ can be substituted in $B_{\varphi'_H}$ in Eqs. (17-19) which results in

$$\begin{aligned} B_{\varphi'_H} &= \alpha' \Delta H_Y + \beta' \Delta H_X \\ &= \beta \Delta H_Y + \alpha \Delta H_X \\ &= B_{90° - \varphi_H} \end{aligned} \quad . \tag{22}$$

As such, we can derive the current induced effective fields using Eqs. (17) and (21) as

$$\Delta H_Y = \frac{\alpha B_{\varphi_H} - \beta B_{90° - \varphi_H}}{\alpha^2 - \beta^2} \quad , \tag{23}$$

$$\Delta H_X = \frac{\beta B_{\varphi_H} - \alpha B_{90° - \varphi_H}}{\beta^2 - \alpha^2} \quad . \tag{24}$$

By using the point symmetry of $\alpha$ and $\beta$ curves at angle of 135°, as seen in Fig. 1 (b-d), a general expression for the coefficients can be obtained as $\alpha(\xi, \varphi_H) = -\beta'(\xi, 270° - \varphi_H)$ and $\beta(\xi, \varphi_H) = -\alpha'(\xi, 270° - \varphi_H)$. The effective fields are then given by

$$\Delta H_Y = \frac{\alpha B_{\varphi_H} + \beta B_{270° - \varphi_H}}{\alpha^2 - \beta^2} , \qquad (25)$$

$$\Delta H_X = \frac{\beta B_{\varphi_H} + \alpha B_{270° - \varphi_H}}{\beta^2 - \alpha^2} . \qquad (26)$$

The symmetry in the curves enable the determination of current induced effective fields to be cross-checked using the same measurement data.

## 2.    Results and Discussion

To investigate the angular dependence of the SOT effective fields, Ta(5)/Pt(3)/Co(0.9)/Pt(1)/Co(0.9)/Ta(1) reference stacks were grown at room temperature on Si/SiO$_2$(300 nm) substrates using DC magnetron sputtering deposition technique. The numbers in brackets correspond to the respective film thicknesses in nm. The magnetization hysteresis loop as obtained using Magneto optical Kerr setup reveals that the continuous film has an intrinsic PMA with a coercivity of 200 Oe. A test sample was grown by depositing a 10 nm BTO layer directly on the reference stack sputtered, using RF magnetron sputtering technique. Before the BTO layer growth, the reference stack was reverse sputtered in Ar plasma for 30 s twice to remove any naturally oxidization of the Ta layer.  No field or heat treatment was carried out to the samples.

Both reference and test sample were patterned into Hall cross structure using a combination electron beam lithography and argon ion milling techniques. The Hall cross structures comprise of a 6-µm-wide current channel and Hall bars. The length of the current channel and the Hall bars are 40 µm. A Keithley 6221 AC current source and a 7265 Dual Phase DSP Lock-In Amplifier were used for harmonic Hall voltage measurement. The AC current frequency was set to 309 Hz at room temperature for all our measurements.

## 2.1. Determination of PHE&SMR to AHE ratio

The ratio of the P&S resistance to AHE resistance, $\xi = \Delta R_{P\&S} / \Delta R_A$, is an intrinsic material property and knowledge of the ratio is crucial for evaluating the coefficient $\alpha$ and $\beta$. Due to limitations in our experimental setup in applying large fields to saturate the magnetization of the PMA wire in-plane, a technique is devised for determining the P&S resistance using low external fields ($H << H_K$). From Eq. (1), the maximum and minimum values for $R_H$ due to PHE or SMR would occur when the azimuthal angle is set to +45° and −45° respectively. The AHE resistance is measured to be ~0.673 Ω with an $I_{ac}$ current in the wire of $1 \times 10^{11}$ A/m². Normalizing the Hall resistance $R_H$ in terms of the AHE resistance $\Delta R_A$ results in

$$\tilde{R}_H = \frac{R_H}{\Delta R_A} = \frac{1}{2}\cos\theta + \frac{1}{2}\xi\sin^2\theta\sin 2\varphi .$$ (27)

Summing or subtracting $\tilde{R}_H$ measured at azimuthal angles of +45° and −45° leads to the following expressions which separates both terms in $\tilde{R}_H$:

$$\Delta\tilde{R}_H^+ = \tilde{R}_H^{+45^\circ} + \tilde{R}_H^{-45^\circ} = \cos\theta ,$$ (28)

$$\Delta\tilde{R}_H^- = \tilde{R}_H^{+45^\circ} - \tilde{R}_H^{-45^\circ} = \xi\sin^2\theta = \xi\left[1 - \left(\Delta\tilde{R}_H^+\right)^2\right].$$ (29)

Shown in Fig. 2 (a) are the Hall resistance curves for sample with magnetization set along +$z$ orientation as the external field is swept at azimuthal angles of ±45°. From Equation (3), the difference between the two curves in Fig. 2 (a), $\Delta \tilde{R}_H^-$, is calculated and plotted as a function of $[1-\left(\Delta \tilde{R}_H^+\right)^2]$, as seen in Fig. 2 (b). A linear relation is observed and the gradient of the slope gives the term $\xi$. For our sample structure, the ratio is calculated to be $\xi = 0.205 \pm 0.001$.

The ratio $\xi$ can be cross-checked by an alternative method by comparing the parabolic coefficients $b_\omega$ (Eq. (15)) from the first harmonic Hall voltage directly.

The parameter $b_\omega = \partial^2 V_\omega / \partial H^2$ can be extracted by a parabolic fit of the measured data points of the first harmonic Hall voltage as a function of the externally applied field, with the externally applied field directed at angles $\pm 45°$ with respect to the current direction. The fitting parameters at applied fields with azimuthal angle $\pm 45°$ can be read as

$$b_\omega^{\pm 45°} = \frac{\partial^2 V_\omega}{\partial H^2(\varphi_H = \pm 45°)} = -\frac{\Delta R_A}{2H_K^2}\left(1 \mp 2\frac{\Delta R_{P\&S}}{\Delta R_A}\right)I_{ac}, \tag{30}$$

An expression for the ratio $\xi$ without prior knowledge of the anomalous Hall resistance $\Delta R_A$ as

$$\xi = \frac{1}{2}\frac{\left(b_\omega^{-45°} - b_\omega^{+45°}\right)}{\left(b_\omega^{-45°} + b_\omega^{+45°}\right)}. \tag{31}$$

This value is obtained by $\xi = 0.203 \pm 0.001$. For all subsequent computations in our experiment, a value of $\xi = 0.204$ is used. This value is obtained from averaging $\xi$ computed by both methods.

## 2.2.    Spin-orbit torque effective fields as function of current

To evaluate the SOT effective fields, first and second harmonic Hall voltages are measured as a function of external in-plane magnetic field for two cases; external field applied along the current direction (x-direction) and applied along the direction transverse to the current (y-direction). The externally applied field is swept in the range of ±1.9 kOe to increase the signal-to-noise ratio. Hall voltages for sample magnetized along ±z orientation is shown in Fig. 3. If the applied field is swept along the x- or y-direction, Eqs. (13) and (14) simplify to

$$V_\omega = \pm \frac{1}{2}\Delta R_A I_{ac} \mp \frac{1}{2}\frac{H^2}{H_K^2}I_{ac}, \quad \text{for } x\text{-}, y\text{-direction}, \tag{32}$$

$$V_{2\omega} = \left(\pm\frac{1}{4}\Delta R_A \Delta H_X - \frac{1}{2}\Delta R_{P\&S}\Delta H_Y\right)\frac{H}{H_K^2}I_{ac} \quad \text{for } x\text{-direction}, \tag{33}$$

$$V_{2\omega} = \left(\pm\frac{1}{4}\Delta R_A \Delta H_Y - \frac{1}{2}\Delta R_{P\&S}\Delta H_X\right)\frac{H}{H_K^2}I_{ac} \quad \text{for } y\text{-direction}. \tag{34}$$

Fig. 3 (a) and (b) show first harmonic Hall voltages for externally applied fields swept along x-direction (a) and along y-direction (b) with a constant current density of $1 \times 10^{11}$ A/m$^2$ in the wire. Figures 3 (a) and (b) exhibit identical curves, as expected from Equation (32). The second harmonic Hall voltages for externally applied fields swept along x-direction and y-direction are shown in Fig. 3 (c) and (d). The second harmonic Hall voltages exhibit linear relationships with external field which is consistent with Eqs. (33) and (34). As seen in Fig. 3 (c), identical slope with an offset voltage at zero applied field of $\Delta V_{2\omega} \sim 0.3$ μV are obtained for magnetization orientations with opposite out-of-plane sign, ±z orientation. The identical slope in Fig. 3 (c) corresponds to a constant coefficient in Eq. (33) as the magnetization is opposed, which give $+\frac{1}{4}\Delta R_A \Delta H_X - \frac{1}{2}\Delta R_{P\&S}\Delta H_Y = -\frac{1}{4}\Delta R_A \Delta H_X - \frac{1}{2}\Delta R_{P\&S}\Delta H_Y$. This implies a constant SOT effective field in y-direction ($\Delta H_Y$) and an SOT effective field with opposite x-direction ($\Delta H_X$), as the magnetization orientation is changed. This is consistent with the SOT effective field description in y-direction of the fieldlike term ($\vec{H}_F = H_F \hat{y}$) and in x-direction of the dampinglike term ($\vec{H}_D = H_D \hat{m} \times \hat{y}$).

In Fig. 3 (d) is shown the second harmonic Hall voltage as the external field is applied along the $y$-direction. The magnitude of slopes for sample with opposite magnetization orientation are identical but exhibit a sign change. This implies from Eq. (34) a sign change of the coefficient, which give $+\frac{1}{4}\Delta R_A \Delta H_Y - \frac{1}{2}\Delta R_{P\&S} \Delta H_X = -(-\frac{1}{4}\Delta R_A \Delta H_Y - \frac{1}{2}\Delta R_{P\&S} \Delta H_X)$. This is consistent with a constant SOT effective field in $y$-direction ($\Delta \vec{H}_Y = \vec{H}_F$) and an SOT effective field with opposite $x$-direction ($\Delta \vec{H}_X = \vec{H}_D$). The same offset voltage at zero applied field of $\Delta V_{2\omega} \sim 0.3$ μV is observed.

From first and second harmonic Hall voltages, the SOT effective fields can be computed using Equations (23) and (24) for external fields applied along $x$-direction ($\varphi_H = 0^o$) and $y$-direction ($\varphi_H = 90^o$). Equations (23) and (24) using annotations for the ratios $B_{\varphi_H = 0^o} = B_X$ and $B_{\varphi_H = 90^o} = B_Y$ simplify to

$$\Delta H_Y = -2 \frac{(B_Y \pm 2\xi B_X)}{1-4\xi^2}, \tag{35}$$

$$\Delta H_X = -2 \frac{(B_X \pm 2\xi B_Y)}{1-4\xi^2}. \tag{36}$$

As expected, Eqs. (35) and (36) reduces to a form consistent with that derived by Hayashi *et al.* [38], where no angular dependent term is present. The respective SOT effective fields obtained using Eqs. (35) and (36) as a function of the current density are shown in Fig. 4.

As seen in Fig. 4 (a), the effective dampinglike field as a function of the current density in the wire shows linear behavior. The effective dampinglike field is obtained to $\sim\pm17.4$ Oe / $10^{11}$ A/m$^2$ for $\pm z$ orientation of the magnetization. In Fig. 4 (b) the effective fieldlike field shows linear behavior at larger current densities with identical slopes for $\pm z$ orientations of the magnetizations. At low current densities, a large signal-to-noise ratio is responsible for the small uncertainty of the fieldlike term. The effective fieldlike field is found to ~8.4 Oe / $10^{11}$ A/m$^2$ for both magnetization orientations. The Oersted field from the Ta/Pt sublayer as well as the Ta top layer is computed according to Hayashi *et al.* [38] which also scales linearly. The Oersted field assumed to originate mainly from the Ta/Pt sublayer, is directed along the negative *y*-direction. The Oersted field corrected effective fieldlike field is obtained to 12.7 Oe / $10^{11}$ A/m$^2$. In a study using a Ta(4)/Pt(3)/Co(0.9)/Ta(1) stack material a dampinglike term of ~80 Oe / $10^{11}$ A/m$^2$ and a fieldlike term of ~50 Oe / $10^{11}$ A/m$^2$ is reported [39]. The increase of almost 5 times might be related to our double layer ferromagnetic layer and a 1.5 nm layer of TaO$_x$ cap used in the reported study. The ratio $\xi$ in our sample is 0.204 whereas in the reported study the ratio is ~0.34, which is significantly larger. The ratio $\Delta H_X / \Delta H_Y$ is ~1.4 for our sample and ~1.6 for the sample reported.

## 2.3. Angular dependence of SOT fields using harmonic Hall voltage scheme in Pt/Co/Pt/Co/Ta/BTO sample

To evaluate SOT effective fields as a function of the azimuthal angle $\varphi_H = \varphi$ with respect to the current direction in PMA structures, first and second harmonic Hall voltages $V_\omega$ and $V_{2\omega}$ are measured as a function of the external applied field with magnetization direction of the FM wire set along the $+z$ orientation for all measurements. The current density was set to $1 \times 10^{11}$ A/m$^2$ in the wire to increase the signal-to-noise ratio and the azimuthal angle $\varphi_H$ is varied in 5° steps. For each field sweep at respective azimuthal angle, the first harmonic Hall voltage is fitted by a parabolic function to obtain the second derivative in $H$, $b_\omega = \partial^2 V_\omega / \partial H^2$, and the second harmonic Hall voltage is fitted by a linear function to obtain the first derivative in $H$, $b_{2\omega} = \partial V_{2\omega} / \partial H$. The fitting parameters $b_\omega$ and $b_{2\omega}$ as a function of the azimuthal angle $\varphi_H$ are shown in Fig. 5 (a) and (b) respectively. As seen in Figure 5 (a), the parameter $b_{2\omega}$ exhibits a predominantly $\cos\varphi_H$ trend. In Figure 5 (b) the parameter $b_\omega$ exhibits a clear $\sin 2\varphi_H$ behavior which is expected from Equation (15). To remove the dependencies of the current and anisotropy field, the ratio $B_{\varphi_H} = b_{2\omega} / b_\omega$ is computed, as shown in Fig. 5 (c). Assuming the SOT effective fields are constant with azimuthal angle $\varphi_H$, the ratio $B_{\varphi_H}$ can be directly fitted using Eq. (16) and a ratio of $\xi = 0.204$ to evaluate mean values for SOT effective fields, as seen in Fig. 5 (c). The mean values for the SOT effective fields from the fitting function in Fig. 5 (c) are determined as 8.4 Oe/$10^{11}$ A/m$^2$ for the fieldlike field and 17.0 Oe / $10^{11}$ A/m$^2$ for the dampinglike field. This is consistent with the SOT effective fields computed in the previous section.

To evaluate the dependency of the SOT effective fields on the azimuthal angle, the SOT effective fields are computed from $B_{\varphi_H}$ using Eqs. (23-26) in the analytical solution section, as seen in Fig. 6. Both SOT effective fields can be cross-checked by using different symmetry axes with the same measured data points. The symmetry axes are indicated as red dashed lines in Fig. 6. At symmetry axes where $\alpha^2 = \beta^2$, no effective field can be computed. At angles, $\varphi_H$, close to the symmetry axes, the computation delivers larger errors whereas small errors are obtained between symmetry axes. The mean value for the dampinglike field is found to ~17.4 Oe / $10^{11}$ A/m$^2$ which is identical to the value obtained from angles 0° and 90°. Comparing Fig. 6 (a) and 6 (b), we obtain similar relations with two minima at azimuthal angles of 45° and 225°. This provides a clear indication of azimuthal angular dependence of the dampinglike effective field in the Pt/Co/Pt/Co/Ta/BTO heterostructure. The mean value of the fieldlike field is found to ~8.4 Oe / $10^{11}$ A/m$^2$ which is similar to the value obtained at 0° and 90°. Comparing Fig. 6 (c) and (d) for the fieldlike field, no such relation is observed.

The lack of angular dependency for the fieldlike term may be attributed to the smaller ratio of the Rashba spin-orbit coupling to the exchange coupling of the magnetic material. As predicted by K. Lee *et al.* [34], when the Rashba spin-orbit coupling is larger than the exchange coupling, an angular dependency of the SOT terms might arise.

## 2.4. Angular dependence of SOT fields in Pt/Co/Pt/Co/Ta test sample

We also performed harmonic angular dependence measurements for the reference sample of Pt/Co/Pt/Co/Ta without BTO capping layer. The Ta top layer should be naturally oxidized which may be responsible for small variations of the SOT effective field magnitudes. The SOT effective fields of the reference sample without BTO as a function of the azimuthal angle $\varphi_H$ is shown in Fig. 7. The dampinglike field for different symmetry axes indicated as dashed lines is shown in Fig. 7 (a) and (b). The mean value of the dampinglike field is found to ~17.4 Oe / $10^{11}$ A/m$^2$ which is slightly larger compared to the test sample with BTO. Comparing Fig. 7 (a) and 7 (b), no significant relationship is observed. In Fig. 7 (c) and (d) are shown the fieldlike fields for different symmetry axes. The fieldlike field is found to ~7.8 Oe / $10^{11}$ A/m$^2$ which is slightly lower compared to the test sample without BTO. Comparing Fig. 7 (c) and (d) indicates no significant relationship. It should be noted, that similarities in Fig. 7 (a) and (c) as well as in Figures 7 (b) and (d) indicate measurement artifacts and are not related to angular dependence of SOT effective fields. The lack of SOT angular dependence on the azimuthal angle in the Pt/Co/Pt/Co/Ta reference sample may be related to the small Rashba spin-orbit coupling as the sample is weakly oxidized.

## 2.5. Effective magnetic anisotropy field dependence on azimuthal angle

From Equation (12) the out-of-plane magnetization vector modulation via current depends on the effective out-of-plane anisotropy field. The effective out-of-plane anisotropy field can be expressed as $H_K = 2K_{Eff}/M_S$ with $K_{Eff}$ the effective magnetic anisotropy constant and $M_S$ the saturation magnetization. As such, to investigate the effect of the out-of-plane anisotropy to the SOT effective fields, the out-of-plane anisotropy field can be extracted from the same harmonic Hall voltage measurement reformulating Eq. (15) in terms of the out-of-plane anisotropy field as

$$H_K = \sqrt{-\frac{1}{2b_\omega}\left(\Delta R_A - 2\Delta R_{P\&S}\ \sin 2\varphi_H\right)I_{ac}} = \sqrt{-\frac{\Delta R_A}{2b_\omega}\left(1 - 2\xi \sin 2\varphi_H\right)I_{ac}}\ . \tag{37}$$

To confirm the validity of this equation, we performed large field sweep at $\varphi_H = 45°$ with maximum field range of $\pm 11.5$ kOe as seen in Fig. 8 (a). The out-of-plane anisotropy field can be extracted from the cross section of the plotted parabolic fit with the nearly saturated Hall resistance, as seen in Fig. 8 (a). The effective out-of-plane anisotropy field is estimated to ~10.5 kOe.

Using the coefficient $b_\omega$ from the parabolic fit for low external fields a precise value can be computed for the out-of-plane anisotropy field using Eq. (37). In Fig. 8 (b) the out-of-plane anisotropy field is shown as a function of the current density in the wire for low field range of $\pm 1.9$. The AHE resistance of 0.673 $\Omega$ and a ratio $\xi$ of 0.204 is used. The out-of-plane anisotropy field was averaged by using the first harmonic Hall voltages measured at externally applied field angles of $\varphi_H = [0°, 90°, 180°, 270°]$, where no effective contributions of the PHE or SMR is expected. The out-of-plane anisotropy effective field decreases linearly from ~10.21 kOe to around ~10.13 kOe as the current density in the wire is increased from 0.1 to $1 \times 10^{11}$ A/m$^2$. This decrease might be related to the increase in SOT effective fields with increasing current density in the wire [30].

The out-of-plane anisotropy field as a function of azimuthal angle is shown in Fig. 8 (c). The out-of-plane anisotropy field varies in a range of 550 Oe with a mean value of ~10.1 kOe. Two maxima at angles $\varphi_H \approx 45°$ or $\varphi_H \approx 270°$ and two minima at angles $\varphi_H \approx 180°$ or $\varphi_H \approx 360°$ with nearly linear relationship between the extrema are observed. The out-of-plane anisotropy field can generally be corrected by the in-plane anisotropy field and an externally applied out-of-plane field component, which could arise from a misalignment in our setup while rotating the sample in the experiment, which is given by $H_{K-corr} = H_K - H\delta\cos(\varphi_H + \chi) - H_I \sin^2 \varphi_H$, where $\delta$ and $\chi$ are the offset polar and azimuthal angles from sample to the applied field direction. $H_I$ is the in-plane anisotropy effective field. As the out-of-plane anisotropy field cannot be fitted to this equation, the cause of the relationship remains unknown. The out-of-plane anisotropy field of the reference sample without BTO as a function of the azimuthal angle is shown in Fig. 8 (d). The out-of-plane anisotropy field varies in a range of less than ~150 Oe at a mean value of around 11.0 kOe. The smaller variation may be related to noise which does not exhibit any significant trend. Still a direct influence of the out-of-plane anisotropy field to the obtained SOT effective fields is not conclusive.

## 3.    Conclusion

In this work, we have developed and experimentally tested a harmonic Hall technique to quantify SOT effective fields as a function of the magnetization azimuthal angle with respect to the current direction in PMA structures. Based on the energy balance equation, analytical solutions are provided to extract the SOT effective fields from first and second harmonic Hall voltages measured at two different azimuthal angles of the external field. Two methods to quantify the ratio of two different symmetries in Hall voltages originated from AHE on one side and from PHE and SMR on the other side are provided for low external magnetic field sweeps. Experimental study was carried out on Pt/Co/Pt/Co/Ta/BTO test sample and Pt/Co/Pt/Co/Ta reference sample with PMA. The ratio of the AHE resistance to PHE&SMR resistance for the test sample has been found to 0.204 using both methods. The dampinglike term for the test sample is found to be ~17.4 Oe per $1 \times 10^{11}$ A/m$^2$ and the fieldlike term to be ~12.7 Oe per $1 \times 10^{11}$ A/m$^2$ corrected by computed Oersted field of ~4.3 Oe per $1 \times 10^{11}$ A/m$^2$. An angular dependence of the dampinglike term is found for the test sample with minima at 45° and 225° whereas no angular dependence is found for the fieldlike term. For the reference sample Pt/Co/Pt/Co/Ta without BTO, no angular dependence of both terms is observed which is in agreement with the assumption of a weakly dependent azimuthal dependence of the SOT effective fields [33, 40]. Thus, we assume the angular dependence of dampinglike field originating from the Rashba effect from an intrinsic electric field in the BTO or from the BTO/Ta interface. We also quantified the out-of-plane anisotropy field as a function of azimuthal angle for both samples. A weakly dependent behavior in a range of 150 Oe with an anisotropy field of ~11.0 kOe is found for the sample without BTO whereas a strong angular dependence of the BTO sample is found with peaks at 45° and 275° with a modulation of ~550 Oe at an anisotropy field ~10.1 kOe. However, the correlation of the anisotropy field and the dampinglike field as a function of the magnetization azimuthal angle remains to be debatable.

# 4.  Acknowledgments

This work was supported by the Singapore National Research Foundation, Prime Minister's Office, under a Competitive Research Programme (Non-volatile Magnetic Logic and Memory Integrated Circuit Devices, NRF-CRP9-2011-01), and an Industry-IHL Partnership Program (NRF2015-IIP001-001). The work was also supported by a MOE-AcRF Tier 2 Grant (MOE 2013-T2-2-017). WSL is a member of the Singapore Spintronics Consortium (SG-SPIN).

# 6.    Figure Captions

**Figure 1.** (a) Schematic of the Hall voltage measurement setup. (b) - (d) Computed coefficients $\alpha$ and $\beta$ as a function of the angle $\varphi_H$ for (b) $\xi = 0$, (c) $\xi = 0.2$ and (d) $\xi = 0.4$.

**Figure 2.** (a) Hall resistance as a function of applied field with azimuthal angle $\pm 45°$. The polar angle $\theta_H$ is set to $\pi/2$ as defined in Fig. 1. (b) The subtracted normalized resistance $\Delta \tilde{R}_H^-$ as a function of $\left[ 1 - \left( \Delta \tilde{R}_H^+ \right)^2 \right]$ and linear fit to the data points.

**Figure 3.** (a), (b) first harmonic Hall voltage as a function of the applied in-plane field. (a) shows the Hall voltage with field pointing along $x$-direction and (b) with field pointing along $y$-direction. (c), (d) second harmonic Hall voltage as a function of the applied in-plane field. (c) shows the Hall voltage with field pointing along $x$-direction and (d) with field pointing along $y$-direction. The black points correspond to a sample magnetized $+z$ orientation and the red points correspond to a sample magnetized in $-z$ orientation.

**Figure 4.** (a) Effective dampinglike field as a function of the applied current density in the wire for sample magnetized along $+z$ orientation. (b) Effective fieldlike field and calculated Oersted field as a function of the current density for sample magnetized along $+z$ orientation. The Oersted field was calculated from currents flowing in the nonmagnetic channels.

**Figure 5.** Sample was magnetized along $+z$ orientation. (a) Fit from the second derivative of the first harmonic Hall voltage as a function of the azimuthal angle $\varphi_H$. (b) Linear fit of the second harmonic Hall voltage as a function of the azimuthal angle $\varphi_H$. (c) $B_{\varphi_H}$ as a function of the azimuthal angle $\varphi_H$ and a fit according to Eq. (16).

**Figure 6.** Computed SOT effective fields for the BTO sample as a function of azimuthal angle $\varphi_H$ of the applied field with respect to current direction. (a) Dampinglike effective field according to Equation (24). (b)

Dampinglike effective field according to Equation (26). (c) Fieldlike effective field according to Equation (23). (d) Fieldlike effective field according to Equation (25). The dashed lines are symmetry axes.

**Figure 7.** Computed SOT effective fields for the reference sample without BTO as a function of azimuthal angle $\varphi_H$ of the applied field with respect to current direction. (a) Dampinglike effective field according to Equation (24). (b) Dampinglike effective field according to Equation (26). (c) Fieldlike effective field according to Equation (23). (d) Fieldlike effective field according to Equation (25). The dashed lines are symmetry axes.

**Figure 8.** (a) Hall resistance of BTO sample as a function of applied magnetic field along 45° in-plane angle with respect to the current. The parabolic fit is carried out taking data points between -3 kOe and 3 kOe for positive Hall resistance into account. (b) Averaged effective anisotropy field, $H_K$, of BTO sample of azimuthal angles, $\varphi_H = [0°, 90°, 180°, 270°]$ as a function of the current density in the wire. The ratio $\xi$ is 0.204. (c) Effective anisotropy field $H_K$ of BTO sample as a function of azimuthal angle $\varphi_H$ computed ratio $\xi$ of 0.204. (d) Effective anisotropy field $H_K$ of sample without BTO as a function of azimuthal angle $\varphi_H$ computed ratio $\xi$ of 0.204. For (a), (c) and (d) the current density in the wire was set to $1 \times 10^{11}$ A/m$^2$.

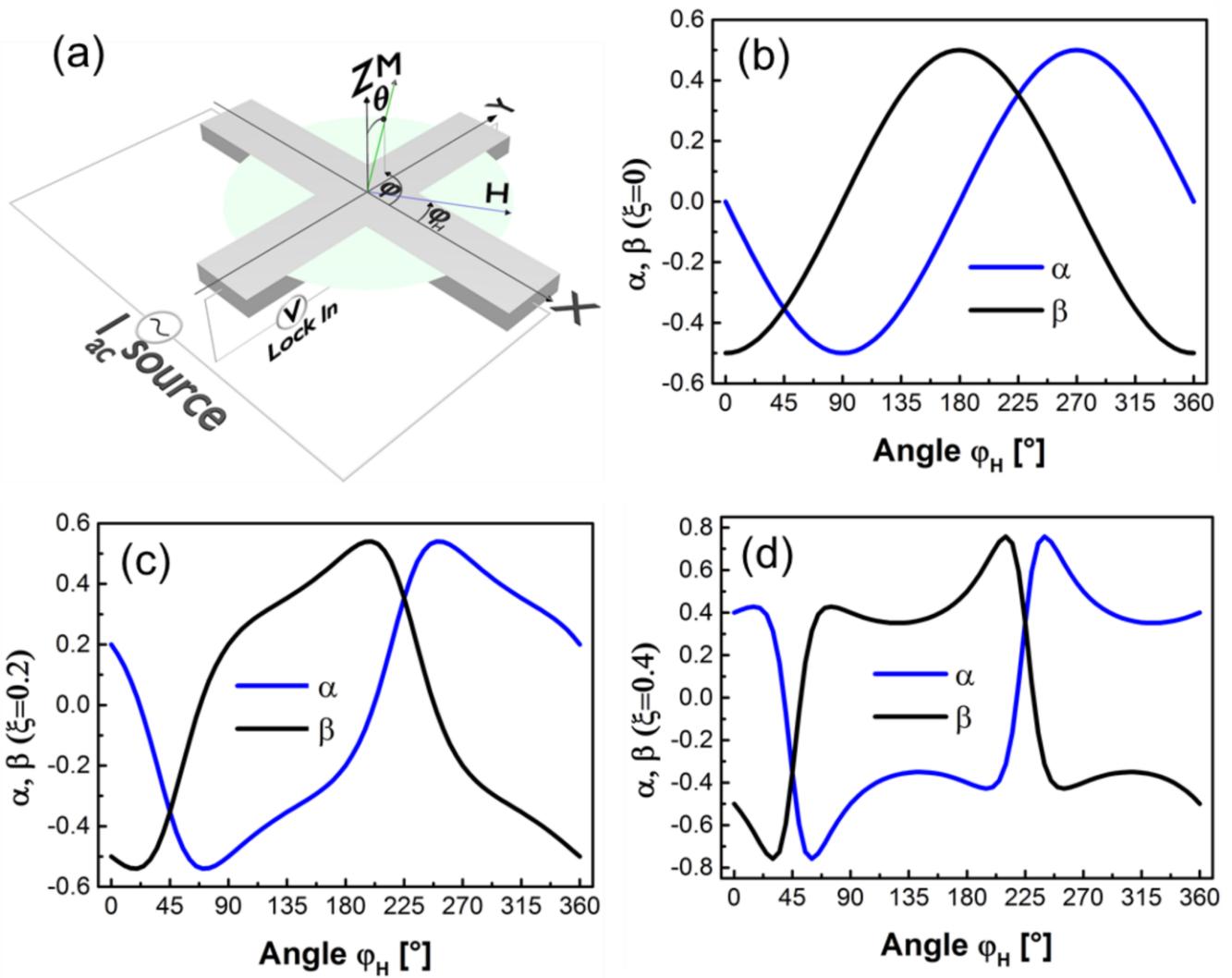

Figure 1, Engel *et al.*

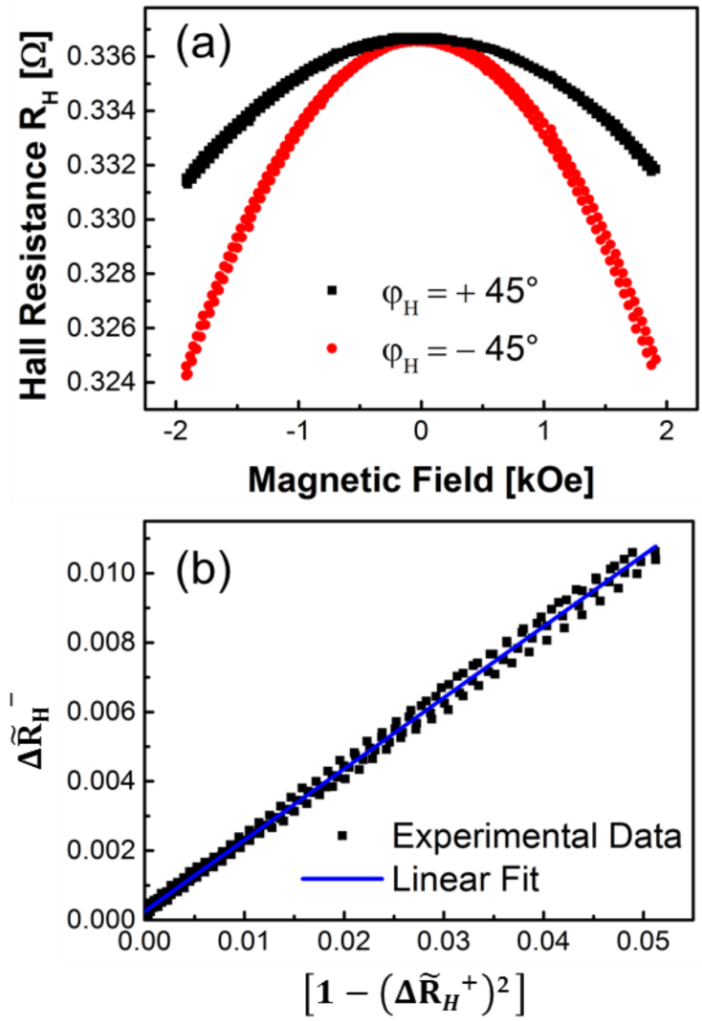

Figure 2, Engel *et al.*

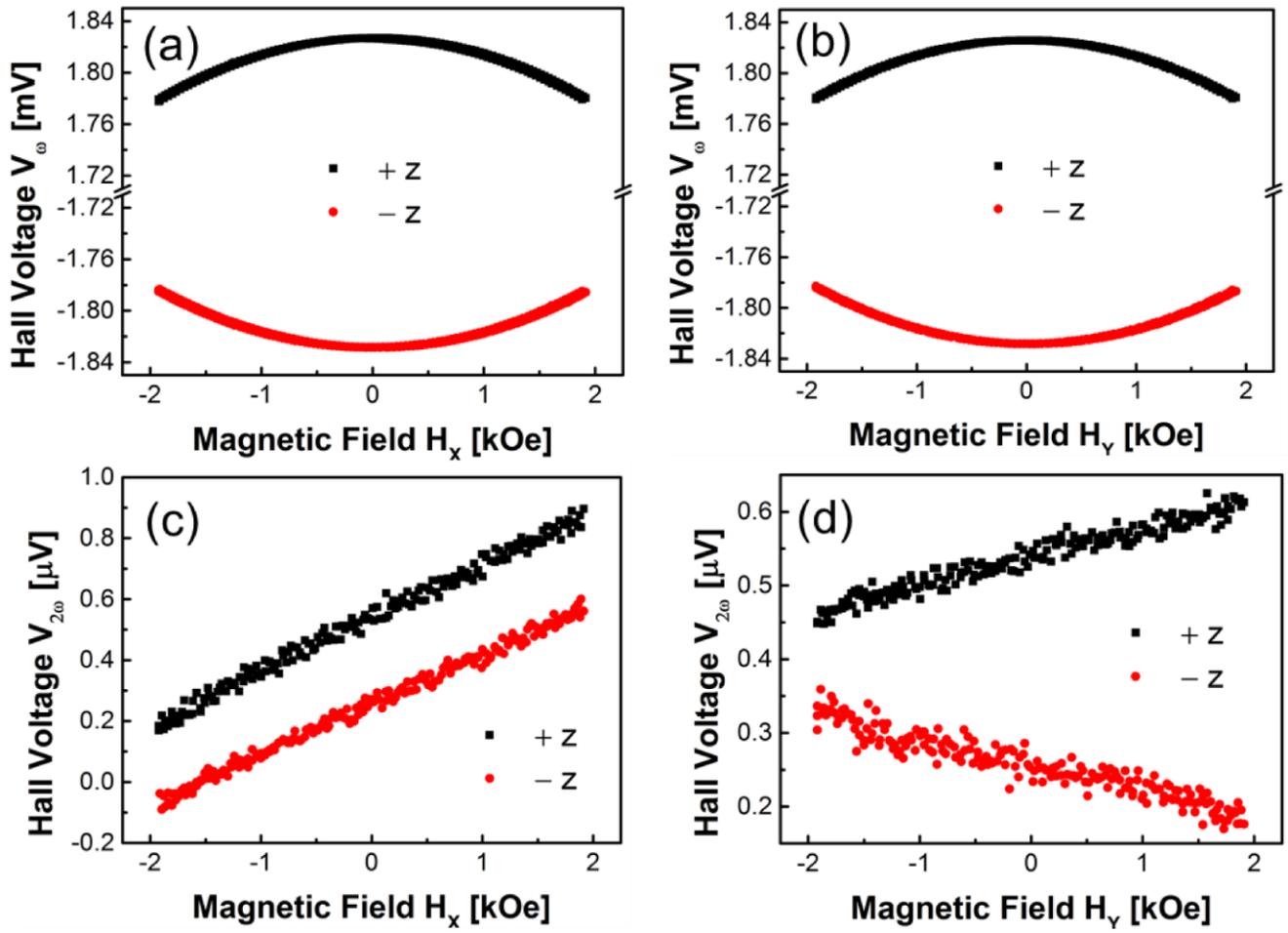

Figure 3, Engel *et al.*

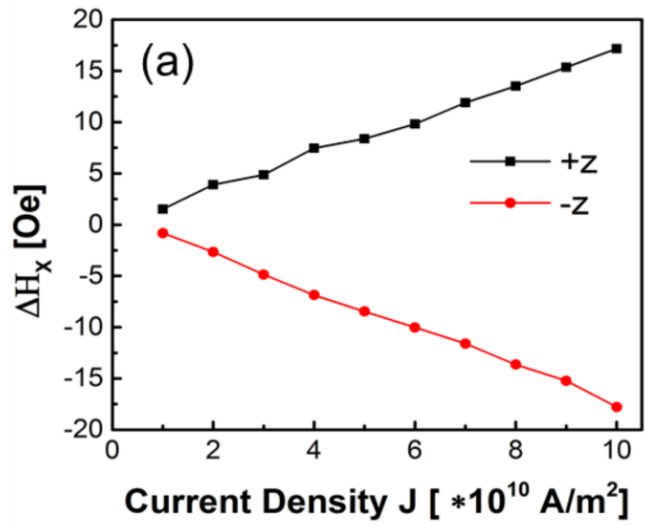

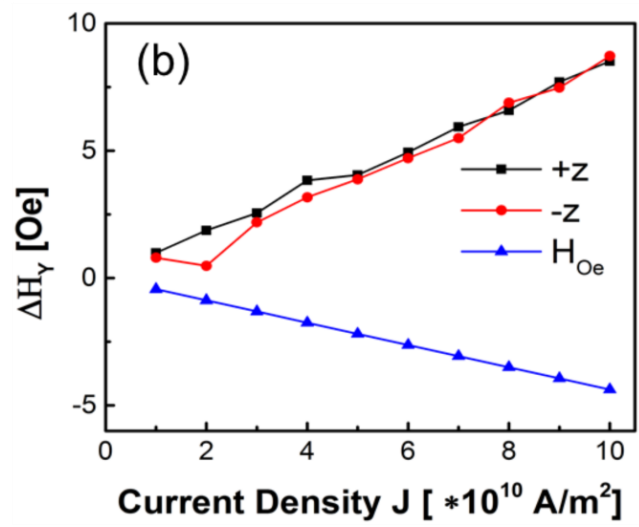

Figure 4, Engel *et al.*

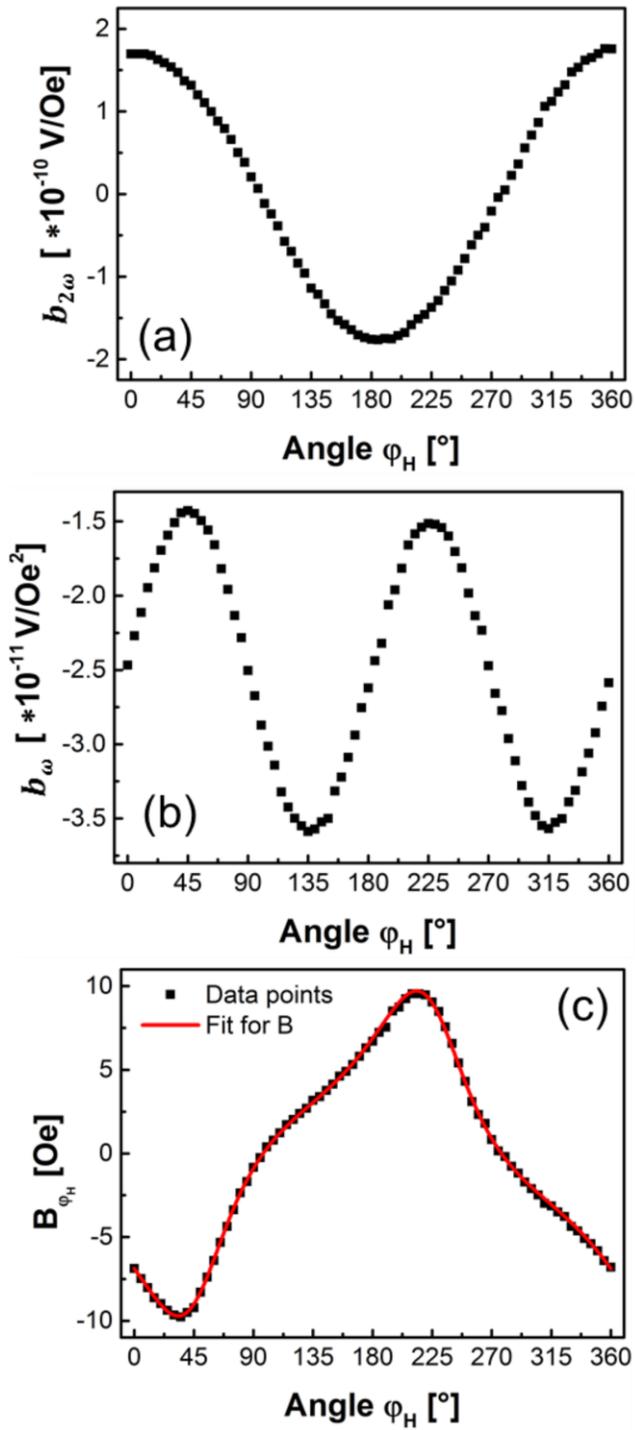

Figure 5, Engel *et al.*

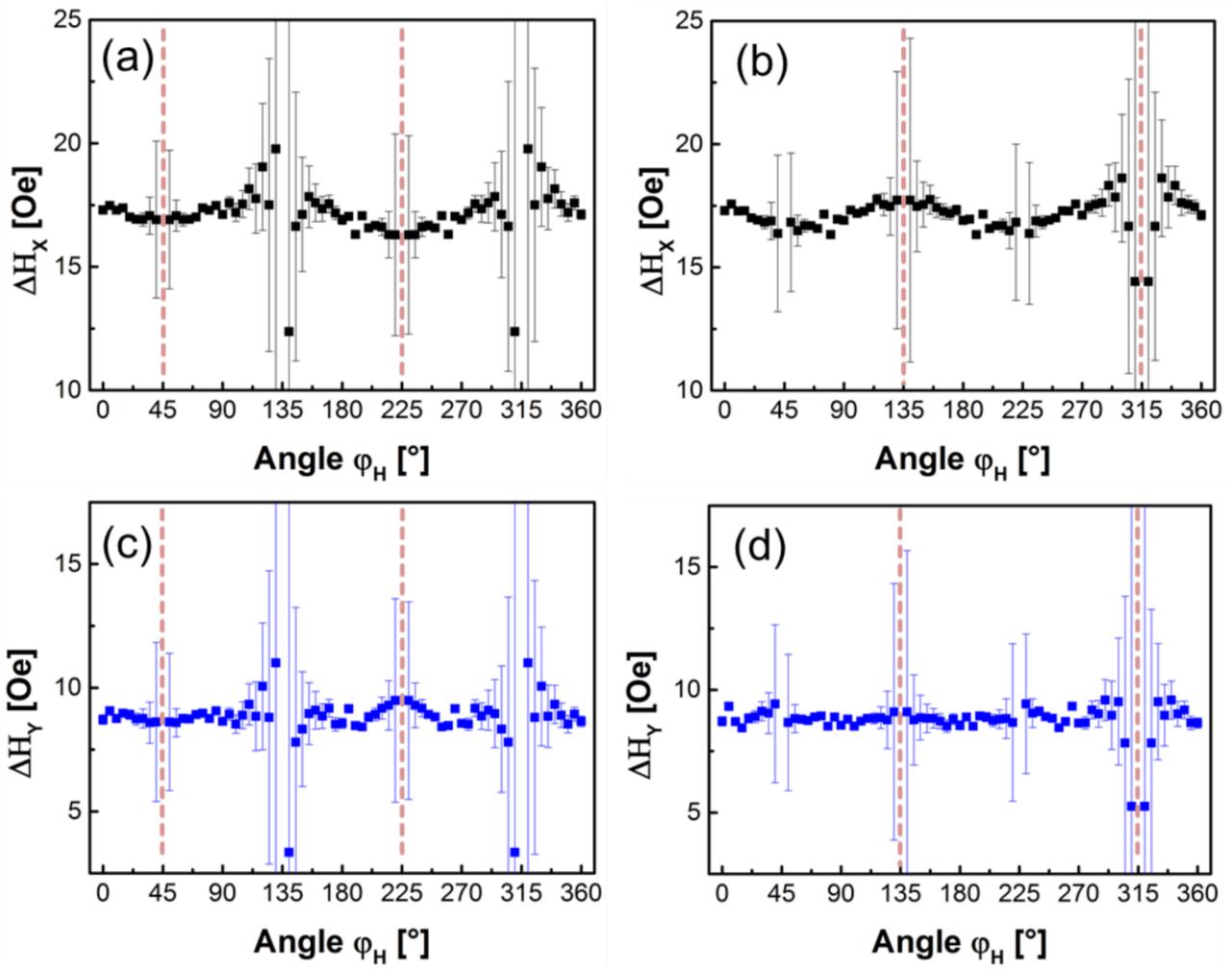

Figure 6, Engel *et al.*

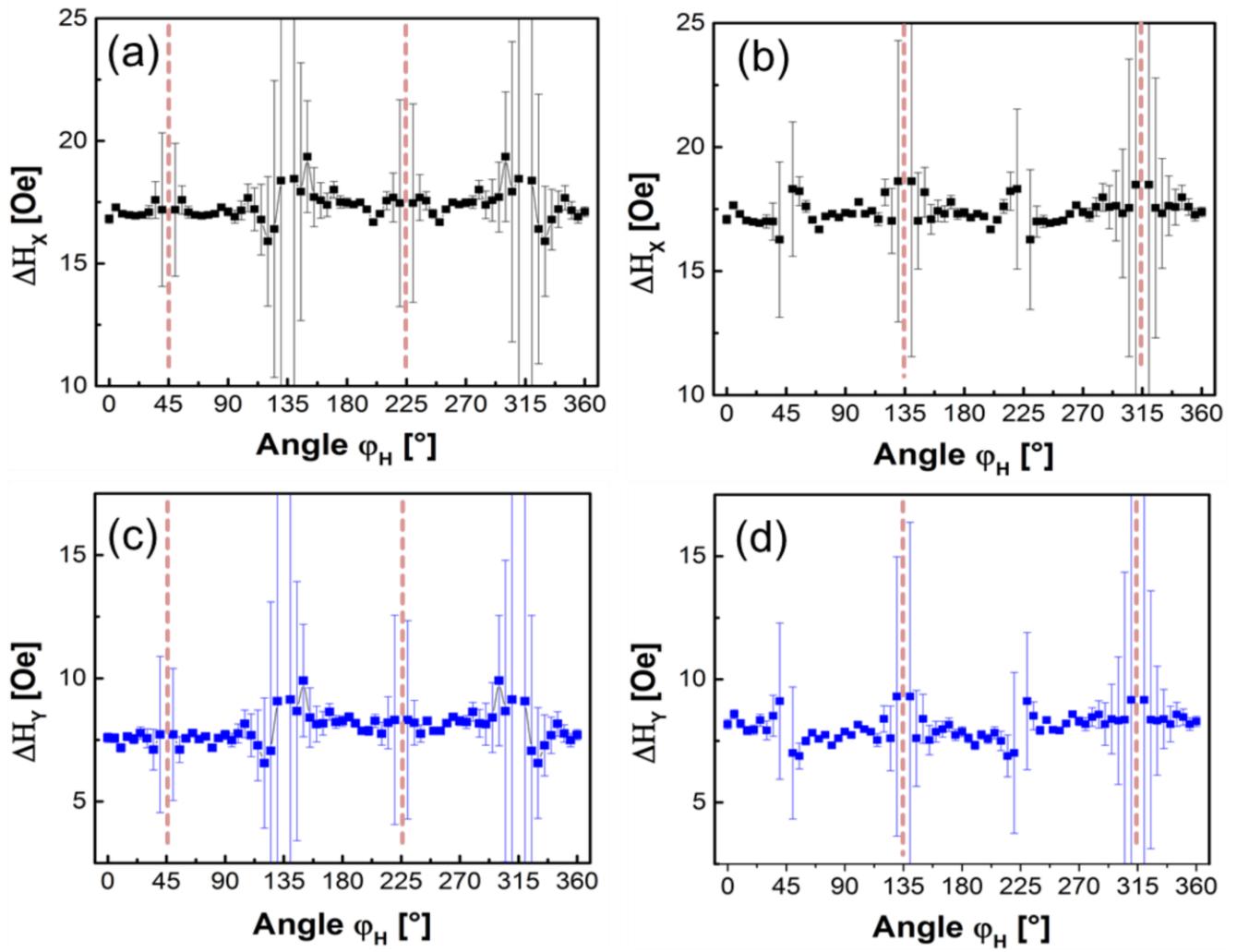

Figure 7, Engel *et al.*

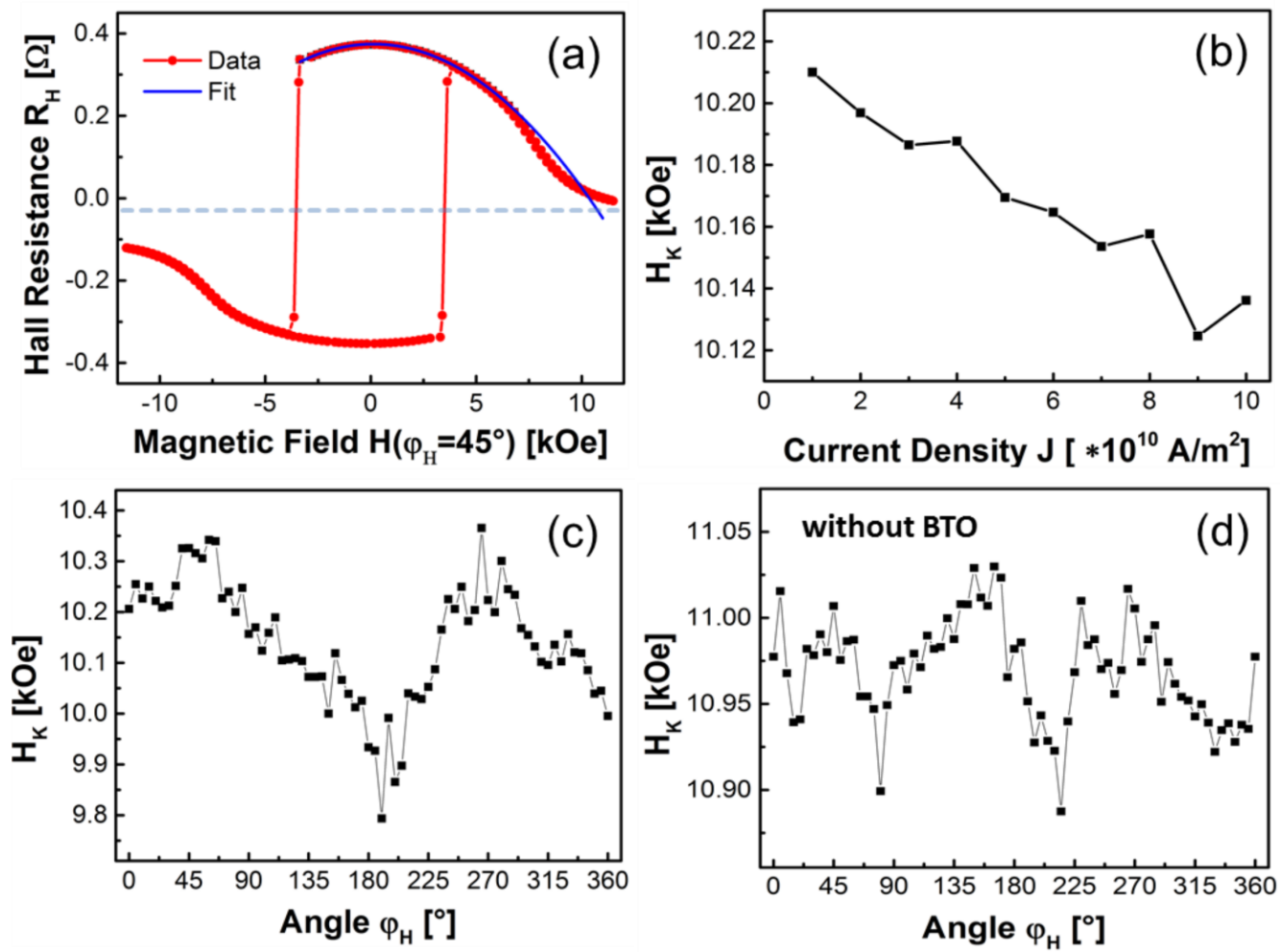

Figure 8, Engel *et al.*